\documentclass[twocolumn]{aastex631}

\usepackage{amsmath}
\usepackage{natbib}
\usepackage{multirow}
\usepackage{array}
\newcolumntype{P}[1]{>{\centering\arraybackslash}p{#1}}

\shorttitle{Identifying the object with mass range in $(2.2-3)M_\odot$}
\shortauthors{Du et al.} 

\begin{document}
\title{How to identify the object with mass range of $(2.2-3)M_\odot$ in the merger of compact star systems}
\author[0000-0002-4375-3737]{Zhao-Wei Du}
\affiliation{Guangxi Key Laboratory for Relativistic Astrophysics, Department of Physics, Guangxi University, Nanning 530004, China; lhj@gxu.edu.cn}

\author[0000-0001-6396-9386]{HouJun L\"{u}} 
\affiliation{Guangxi Key Laboratory for Relativistic Astrophysics, Department of Physics, Guangxi University, Nanning 530004, China; lhj@gxu.edu.cn}

\author{Xiaoxuan Liu} 
\affiliation{Guangxi Key Laboratory for Relativistic Astrophysics, Department of Physics, Guangxi University, Nanning 530004, China; lhj@gxu.edu.cn}

\author[0000-0002-8174-0128]{XiLong Fan} 
\affiliation{School of Physics Science and Technology, Wuhan University, No.299 Bayi Road, Wuhan, Hubei, China}

\author[0000-0002-7044-733X]{EnWei Liang}
\affiliation{Guangxi Key Laboratory for Relativistic Astrophysics, Department of Physics, Guangxi University, Nanning 530004, China; lhj@gxu.edu.cn}

\begin{abstract}
High-frequency gravitational-wave (GW) radiation has been detected by LIGO-Virgo-KAGRA in the merger of compact stars. However, two GW events, GW190814 and GW200210, the mass of one companion object falls into the mass region of $(2.2-3)\rm~M_\odot$, and how to identify such object (e.g., as a low-mass black hole (BH) or a massive neutron star (NS)) remains an open question. In this paper, we propose a method to identify the mystery compact object (MCO) with the mass region of $(2.2-3)\rm~M_\odot$ in a binary system via the possible electromagnetic (EM) radiations before and after the mergers. A multi-band EM emission can be produced with $L\propto(-t)^{7/4}$ (or $L\propto(-t)^{-5/4}$) during the inspiral phase due to the BH battery (or interaction magnetospheres) mechanism, and a bright (or dark) kilonova emission is powered by radioactive decay with ejecta mass ratio $q>1.7$ (or $q<1.7$) during the post-merge state when MCO is as a low-mass BH (or massive NS) to merger with NS. Moreover, by considering the merger system between MCO and a BH when MCO is a massive NS, we find that it requires the BH with high spin (e.g., $a\sim0.8-0.99$) to make sure the tidal disruption event (TDE) occurred, and a multi-band precursor emission and bright kilonova emission can also be produced during the inspiral phase and post-merge state, respectively. In any case, no matter which mechanism we adopt, such precursor emissions are too weak to be detected by most current telescopes unless the distance is close enough.

\end{abstract}

\section{Introduction} \label{sec:intro}
The coalescence of two compact stars is expected to produce both gravitational wave (GW) emission and electromagnetic radiation (\citeauthor{2014ARA&A..52...43B} \citeyear{2014ARA&A..52...43B} for a review). The first direct detection of GW emission, GW150914, which originated from the merger of a binary black hole (BH-BH), was achieved by the Laser Interferometer Gravitational Wave Observatory(LIGO; \citealt{2016PhRvL.116f1102A}). It opened a new window to understand our Universe and marked a milestone in GW astronomy. However, the EM counterparts associated with the merger of a binary black hole have not been detected so far \citep{2018pgrb.book.....Z}, and whether or not the merger of a binary black hole can be accompanied by EM counterparts is still highly debated \citep{2016ApJ...826L...6C,2016ApJ...827L..31Z}.

Fortunately, an EM counterpart associated with a GW event was detected on August 17th, 2017, and it is identified as a binary neutron star (NS-NS) merger event known as GW170817 \citep{2017PhRvL.119p1101A}. The EM counterparts associated with GW170817 include short-duration gamma-ray burst (GRB) and an optical/infrared transient known as kilonova AT2107gfo \citep{2017ApJ...848L..14G, 2017ApJ...848L..15S,2017Natur.551...67P,2017ApJ...848L..29D,2017SciBu..62.1433H,2018ApJ...852L..30P,2018NatCo...9..447Z,2019LRR....23....1M}. This event opened a new window for the study of multi-messenger astronomy and enormously enhanced our understanding of neutron star physics and mergers of binary neutron star systems \citep{2017ApJ...850L..34B, 2017ApJ...850L..19M, 2018ApJ...852L..29R, 2019MNRAS.486.4479L, 2020ApJ...893..146A}. The simultaneously observed GW and EM radiations from the merger of binary neutron stars can help us constrain the Hubble constant \citep{2019ApJ...871L..13F, 2020MNRAS.492.3803H}, identify the origin of the nucleosynthesis of heavy elements \citep{2017Sci...358.1570D, 2017Natur.551...80K,2024MNRAS.529.1154C}, as well as confirm the origin of some short GRBs at least from the coalescence of binary neutron stars \citep{2017ApJ...835..181L,2017ApJ...848L..14G,2018NatCo...9..447Z}.

Besides the mergers of BH-BH and NS-NS, the GW signal powered by the black hole-neutron star (BH-NS) merger is also within the sensitivity of LIGO-Virgo-KAGRA (LVK) detectors \citep{2021ApJ...915L...5A}. Moreover, an EM counterpart is also expected to be produced when a tidal disruption event (TDE) happens or a sufficiently strong magnetic field exists in the neutron star \citep{2011ApJ...742...90M, 2012ApJ...757L...3L, 2022MNRAS.514.5385N,2022ApJ...927...56D}. From the observational point of view, several GW events from the merger of BH-NS have been reported, such as GW200105 \citep{2021ApJ...915L...5A}, GW200115 \citep{2021ApJ...915L...5A}, and GW230529 \citep{2024ApJ...970L..34A}. However, no EM counterpart has been detected so far, and the reason for the lack of EM counterpart in GW200105 and GW200115 is that the dimensionless spin is below the threshold required for TDE to occur in those systems \citep{2020MNRAS.497..726G,2021NatAs...5...46A,2021ApJ...923...66A, 2023PhRvD.108h3018Y}, while GW230529 is attributed to a significant localization error \citep{2024ApJ...970L..34A}. 

Despite the growing number of detected GW events from the merger of compact stars, several GW events, such as GW190814 with $m_1=23.3^{+1.4}_{-1.4}\rm ~M_\odot$ (black hole) and $m_2=2.6^{+0.1}_{-0.1}\rm ~M_\odot$ object \citep{2020ApJ...896L..44A}, GW200210 with $m_1=24.1^{+7.5}_{-4.6}\rm ~M_\odot$ (black hole) and $m_2=2.83^{+0.47}_{-0.42}\rm ~M_\odot$ object \citep{2020ApJ...896L..44A}, and GW 190425 with $m_1=2.1^{+0.5}_{-0.4}\rm~M_\odot$ and $m_2=1.3^{+0.3}_{-0.2}\rm ~M_\odot$ (neutron star) object \citep{2020ApJ...892L...3A}, the median mass of one companion object in binary compact stars of those events fall into the mass region of $(2-3)\rm~M_\odot$. Another example of GW enevt, GW 230529 with $m_1=3.6^{+0.8}_{-1.2}\rm~M_\odot$ and $m_2=1.4^{+0.6}_{-0.2}\rm ~M_\odot$ (neutron star) object, the median mass with 90$\%$ confidence level of one companion object fall into the mass region of $(2-3)\rm~M_\odot$. To identify the object with a mass of $m_2=2.6^{+0.1}_{-0.1}\rm ~M_\odot$ for GW190814, a mass of $m_2=2.83^{+0.47}_{-0.42}\rm ~M_\odot$ for GW200210, a mass of $m_1=3.6^{+0.8}_{-1.2}\rm~M_\odot$ for GW230529, and a mass of $m_1=2.1^{+0.5}_{-0.4}\rm~M_\odot$ for GW190425, remain unknown. However, it is found that all of them exceed the typical range of Tolman-Oppenheimer-Volkoff mass ($M_{\rm TOV}$) of a neutron star, and also exceed the most massive observed neutron star currently known in the pulsar system PSR J0952-0607 ($2.35\pm0.16\rm~M_\odot$) \citep{2022ApJ...934L..17R}. However, due to poor known the equation of state of NS, the maximum mass of NS may reach to a higher mass \citep{1994ApJ...424..823C,2000ApJ...537..351B, 2001ApJ...554..322C, 2010PhRvC..82b5804R, 2016MNRAS.459..646B, 2021ApJ...922..149D, 2022MNRAS.515.3539K, 2024ApJ...962...61M, 2024PhRvD.109b3027Z}. In that case, whether it is a massive neutron star or a low-mass black hole, remains an open question and under debate \citep{2020ApJ...896L..44A,  2021MNRAS.505.1600B, 2021ApJ...908L...1T, 2022PhRvD.105f4063C,2020ApJ...904...80E, 2022ApJ...931..108F, 2022ApJ...936...41L}. Therefore, identifying such an object as a low-mass black hole or a massive neutron star is very important for understanding the formation channel of a low-mass black hole or constraining the equation of state (EOS) of a neutron star due to the uncertainty of the $M_{\rm TOV}$ of a neutron star. 

In this paper, we propose a method to identify the object with a mass region of $(2.2-3)\rm~M_\odot$ in a binary system via the possible EM radiation before and after the merger. In Sec \ref{sec:mass}, we present the details of the reasons for focusing only on the object with the mass region of $(2.2-3)\rm~M_\odot$. Then, we attempt to identify the unknown object with the mass region of $(2.2-3)\rm~M_\odot$ which merges with a neutron star (in Sec \ref{sec:withNS}) or a black hole (in Sec \ref{sec:withBH}). Conclusions and a brief discussion are presented in Sec \ref{sec:end}. Throughout the paper, we adopt CGS units to do the calculations, and some constants are presented as $G=6.67\times10^{-8}~\mathrm{dyn}~\mathrm{g}^{-2}~\mathrm{cm}^{-2}$ and $c=3\times10^{10}~\mathrm{cm}~\mathrm{s}^{-1}$. 

\section{Criterion for selecting mass range}\label{sec:mass}
The selected mass range of the object in a binary compact star, such as a GW190814-like event, is based on the observational and theoretical perspectives on neutron star mass. (1) The lower limit of the selected mass range is obtained from the observed maximum mass of pulsar, PSR J0952-0607, along with black widow and redback pulsars, yielding a $M_\mathrm{max}>2.19M_\odot$ with $1\sigma$ confidence \citep{2022ApJ...934L..17R}. (2) The upper limit of the selected mass range is fixed as $3 M_\odot$ which is derived from the theoretical prediction of the maximum mass of a neutron star by adopting a different equation of state \citep{1974PhRvL..32..324R,1996ApJ...470L..61K}. {Based on the above two reasons, we set a $(2.2-3)M_\odot$ as the mass range for the compact object of interest which is named as Mystery Compact Object (MCO) and discussed in this paper.

One needs to note that it is difficult to identify MCO (i.e., neutron star or black hole) in a binary compact star only based on the GW radiation because it is the existence of degeneracy between a low-mass black hole and a massive neutron star due to some mechanisms that can temporarily support the extra mass of a neutron star before collapsing into a black hole. These includes spin \citep{1994ApJ...424..823C,2016MNRAS.459..646B, 2022MNRAS.515.3539K, 2024ApJ...962...61M} and magnetic field of neutron stars \citep{2000ApJ...537..351B, 2001ApJ...554..322C, 2021ApJ...922..149D, 2010PhRvC..82b5804R, 2024PhRvD.109b3027Z}. So, it is worth studying and identifying the MCO is either a massive neutron star or a low-mass black hole via the possible EM radiation.

\section{MCO merger with a neutron star} \label{sec:withNS}
From the theoretical point of view, whether the MCO is a low-mass black hole or a massive neutron star, there is a high probability that it will produce the EM radiation. The merger of binary neutron stars to produce EM signals associated with GW event was discovered by GW170817/GRB170817A/AT2017gfo \citep{2017ApJ...848L..12A}. On the other hand, the EM signals produced by the merger of BH and NS depend on whether the matter disperses out of the innermost stable circular orbit (ISCO) of BH. The smaller the mass of a BH, the more likely it is to form an accretion disk and release dynamical ejecta because the radius of ISCO of a low-mass BH is also becomes smaller. If this is the case, no matter whether MCO is a low-mass BH or a massive NS, it can produce EM signals (e.g., short GRB and kilonova) when it is merges with the NS \citep{2016ApJ...825...52K,2017CQGra..34o4001F,2019A&A...625A.152B,2018PhRvD..98l3017R}. In this section, we focus on discussing the merger process between MCO and NS, and attempt to present the possible EM radiation to distinguish between those two systems. 

\subsection{MCO as a low-mass black hole scenario}
In this section, we will discuss the physical process and possible EM radiation when MCO is assumed to be a low-mass BH which is merging with an NS. Within the scenario of a low-mass BH-NS system, it roughly contains three phases, namely inspiral, merger, and ring-down. 

\textbf{(1). Inspiral phase:} the relative motion between the NS and the BH, along with their rotations, will generate an electromotive force (emf) when BH enters the light cylinder of a neutron star and is immersed into its dipole magnetic field. Therefore, the plasma in the magnetosphere of the NS could be accelerated by the emf moving along the field lines due to the strong magnetic field, and it is accompanied by the establishment of a current between the BH and the NS. If this is the case, the power absorbed by the intrinsic resistance of BH is transferred to the NS, and the Poynting flux can be emitted which is used to produce the EM radiation. This process is called 'black hole battery mechanism', and the electromotive force of BH can be expressed as \citep{2011ApJ...742...90M, 2016PhRvD..94b3001D},
\begin{equation}
V_{\mathcal{H}}=2R_H\left[\frac{r(\Omega_{\mathrm{orb}}-\Omega_{\mathrm{NS}})}c+\frac a{4\sqrt{2}}\right]B_{\mathrm{NS}}\left(\frac{R_{\mathrm{NS}}}r\right)^3\label{V}
\end{equation}
where $R_\mathrm{H}$ is the horizon radius of BH.  $\Omega_\mathrm{orb}=\sqrt{G(M_\mathrm{BH}+M_\mathrm{NS})/r^3}$ and $\Omega_\mathrm{NS}$ are the angular frequencies of orbit and NS, respectively. $a$ is a dimensionless spin parameter of BH which is defined as the ratio between angular momentum and square of mass ($a=J_\mathrm{BH}/M_\mathrm{BH}^2$ ranging from 0 to 1). Both $a$ and $\Omega_\mathrm{NS}$ are positive (or negative) if they are aligned (or anti-aligned) with $\Omega_\mathrm{orb}$. $B_\mathrm{NS}$ and $R_\mathrm{NS}$ are the strength of the surface magnetic field and the radius of the NS, respectively. The coefficient of 2 is the contribution from both hemispheres due to the spherical symmetry. $r$ is the separation radius between NS and BH, and we adopt the formula proposed by \cite{1964PhRv..136.1224P},
\begin{equation}
    r(t)=\left[\frac{256}5\frac{G^3}{c^5}M_\mathrm{NS}M_\mathrm{BH}(M_\mathrm{BH}+M_\mathrm{NS})(-t)\right]^{1/4}.\label{r}
\end{equation}
Here, $M_\mathrm{BH}$ and $M_\mathrm{NS}$ are the mass of BH and NS, respectively. $t=0$ is the merger time, therefore, $-t$ is the time before coalescence. 

Based on Ohm's law, the power (or luminosity) released by such a circuit is given by 
\begin{equation}
    L_\mathrm{BH-NS}(t)=\frac{V_{\mathcal{H}}^2(t)}{(\mathcal{R}_{\mathcal{H}}+\mathcal{R}_{\mathrm{NS}})^2}\mathcal{R}_{\mathrm{NS}},\label{power}
\end{equation}
where $\mathcal{R}_{\mathcal{H}}=4\pi/c$ is the resistance across the horizon of the black hole, and $\mathcal{R}_{\mathrm{NS}}$ is the effective resistance of the NS and its magnetosphere. Numerical simulation is required to solve the value of $\mathcal{R}_{\mathrm{NS}}$ which is extremely complicated. So that, we adopt $\mathcal{R}_{\mathcal{H}}=\mathcal{R}_{\mathrm{NS}}$ In our calculations \citep{2011ApJ...742...90M}.

Combining the above equations, one can roughly obtain the relationship between battery luminosity and time, namely, $L_\mathrm{BH-NS} \propto (-t)^{-7/4}$. However, the relation is not favored with the occurrence of coalescence or even TDE, namely, $r\geq R_{\rm TDE}=R_\mathrm{NS}(3M_\mathrm{BH}/M_\mathrm{NS})^{1/3}$, where $R_\mathrm{TDE}$ is the separation radius at which tidal disruption occurs. In our calculation, we neglect the effect of the interaction between magnetosphere and radiation, which is notoriously difficult to point out except in the numerical simulations. For example, one can calculate the maximum luminosity of EM radiations $L_{\rm BH-NS, max}=(1.43-1.58)\times10^{43}~\mathrm{erg}~\mathrm{s}^{-1}$ by assuming $M_{\rm MCO}=M_{\rm BH}=(2.2-3)~M_\odot$, $M_{\rm NS}=1.4M_\odot$, $a=0$, $B_\mathrm{NS}=10^{12}~\mathrm{G}$, and $R_\mathrm{NS}=10^6~\mathrm{cm}$.

If this is the case, the multi-band EM signals (e.g., $\gamma$-ray, X-ray, optical, and radio) may be produced by the black hole battery mechanism \citep{2015ApJ...814L..20M, 2021PhRvD.104f3004C}. Figure \ref{fig:1} shows the predicted spectrum of multi-band EM radiations from the black hole battery mechanism by assuming a pure curvature radiation and luminosity distance of 250 Mpc referenced from GW190814. Unfortunately, by comparing the sensitivity of current operating detectors with the results of predictions, it is found that the radiated EM signals are too weak to be detected by Fermi/GBM, Swift/BAT, Swift/XRT, Vera C. Rubin, and FAST. However, the high-energy counterparts (e.g., MeV-GeV emissions) can possibly be detected by Fermi/LAT. 

\textbf{(2). Merger and post-merger phases:} the separation radius between BH and NS is close to the radius of TDE when they loses the angular momentum due to gravitational wave radiation, and the TDE process between BH and NS may occur. One criterion is adopted to determine whether a TDE can occur based on the semi-analytic model proposed by \cite{2018PhRvD..98h1501F},
\begin{equation}
    \frac{M_\mathrm{out}}{M_\mathrm{NS}^\mathrm{b}}=\left[\mathrm{max}\left(\alpha\frac{1-2C_\mathrm{NS}}{\eta^{1/3}}-\beta\hat{R}_\mathrm{ISCO}\frac{C_\mathrm{NS}}{\eta}+\gamma,0\right)\right]^\delta,\label{out}
\end{equation}
where $M_\mathrm{out}$ is the total mass outside the ISCO, and $C_\mathrm{NS}=GM_{\rm NS}/c^2R_{\rm NS}$ is the compactness of the neutron star. $\eta=q/(1+q)^2$ with $q=M_\mathrm{BH}/M_\mathrm{NS}$.  $\alpha=0.406,~\beta=0.139,~\gamma=0.255,~\delta=1.761$ are fixed constants which are taken from \cite{2018PhRvD..98h1501F}. $M^b_\mathrm{NS}$ is the baryonic mass of the neutron star which is calculated according to \citep{2001ApJ...550..426L}
\begin{equation}
    M^b_\mathrm{NS} = M_\mathrm{NS}\Big(1+\frac{0.6C_\mathrm{NS}}{1-0.5C_\mathrm{NS}}\Big).
\end{equation}
$\hat{R}_\mathrm{ISCO}$ is the normalized radius of ISCO of a black hole, and it can be expressed as \citep{1972ApJ...178..347B}
\begin{equation}
\begin{aligned}
    \hat{R}_\mathrm{ISCO} &= \\
    3&+Z_2-\mathrm{sgn}(a)\sqrt{(3-Z_1))(3+Z_1+2Z_2)}.
\end{aligned}
\end{equation}
Here, $Z_1=1+(1-a^2)^{1/3}[(1+a)^{1/3}+(1-a)^{1/3}]$ and $Z_2=\sqrt{3a^2+Z_1^2}$.

In order to produce the EM radiations during the merger, the requirement is that the NS should be tidal disrupted by the low-mass BH. One question is what condition is required for the above requirement? Due to the highly uncertain EoS of NS, we considered the mass of the NS at fixed values of $M_{\rm NS}=1.4M_\odot$, $1.8M_\odot$, and $2.1M_\odot$, and set the compactness of NS as a free parameter. Then, by assuming that the angular momentum of BH is always aligned with the angular momentum of the orbit, we plot the contours for $M_{\rm BH}$, $C_\mathrm{NS}$, and $a$ in Figure \ref{fig:2}. It is found that the probability of tidal disruption occurring within the mass range is relatively high, and it means that the EM signals can be produced during such a merger process. If this is the case, $\gamma-$ray radiation, such as short-duration GRB and its afterglow emission within a small opening angle, may be produced, but it requires that the jet is directed toward the observer \citep{2011ApJ...732L...6R, 2016ApJ...827..102T,2018ApJ...857..128J}. On the other hand, another optical/infrared transient known as kilonova can be generated from ejected material and powered by radioactive decay after near-isotropic r-process \citep{1998ApJ...507L..59L, 2010MNRAS.406.2650M, 2011ApJ...732L...6R, 2013PhRvD..87b4001H, 2013MNRAS.430.2585R, 2014ARA&A..52...43B, 2016NatCo...712898J, 2017ApJ...837...50G, 2021ApJ...912...14Y, 2022ApJ...931L..23L,2023Univ....9..245T}. Here, we focus on the discussion of the kilonova emission which is easier to be detected compared with the collimation of short-duration GRB. 

The main power of kilonova emission from the merger of BH and NS is the radioactive decay of ejected material from the r-process, and the ejected materials include dynamical ejecta and wind ejecta due to disk activity based on different physical properties and formation mechanisms \citep{2020PhR...886....1N}. For dynamical ejecta, it includes ejecta in the direction of orbital angular momentum generated by shock acceleration at the collision surface and ejecta in the direction of the equatorial plane formed by tidal interaction. Here, we adopt the semi-analytic model which is proposed by \cite{2016ApJ...825...52K},
\begin{equation}
\begin{aligned}
    \frac{M_{\mathrm{dyn}}}{M_{\mathrm{NS}}^{\mathrm{b}}}=\mathrm{Max}\Big\{a_1q^{n_1}\frac{1-2C_{\mathrm{NS}}}{C_{\mathrm{NS}}}-a_2q^{n_2}\hat{R}_{\mathrm{ISCO}}&\\
    +a_3\left(1-\frac{M_{\mathrm{NS}}}{M_{\mathrm{NS}}^{\mathrm{b}}}\right)&+a_4,0 \Big\}. \label{dyn}
\end{aligned}
\end{equation}
The coefficients are fixed as constant with $a_1=4.464\times10^{-2}$, $a_2=2.269\times10^{-3}$, $a_3=2.431$, $a_4=-0.4159$, $n_1=0.2497$, and $n_2=1.352$. The material from the merger phase will form an accretion disk around the BH, and the disk which is gravitationally bound does not directly contribute to the kilonova. However, a fraction of material can be ejected through the accretion disk driven by viscosity and neutrino heating. For wind ejecta, the complete formula of wind ejecta is too complex to express analytically, therefore a proportional relationship is adopted to estimate the unbound mass, $M_\mathrm{wind}=f_\mathrm{wind}M_\mathrm{disk}=f_\mathrm{wind} (M_{\rm out}-M_{\rm dyn})$, where $f_\mathrm{wind}$ is a numerical factor ranging from 0 to 0.4 \citep{2020ApJ...889..171K}. So, the total ejecta mass should be the sum of dynamical ejecta mass and wind ejecta mass, i.e., $M_{\rm ej}=M_{\rm dyn}+M_{\rm wind}$.

One can roughly calculate the kilonova emission based on the r-process of ejecta. For example, Figure \ref{fig:3} shows the calculated kilonova emission in the $g$-band by adopting $f=(0-0.4)$, velocity $\beta=0.2$, opacity $\kappa=1\rm~g~cm^{-2}$, and $M_{\rm ej}=(0.013-0.022) ~M_\odot$ which is calculated by Eqs. (\ref{out}) and (\ref{dyn}) with $a=0.3$, $M_{\rm NS}=1.8~M_\odot$, $R_{\rm NS}=12.12~{\rm km}$ \citep{2021ApJ...912...14Y}.

\textbf{(3). Possible fast radio burst emission:} another possible EM signal in the radio band, so called fast radio burst (FRB), may be produced in the binary system of low-mass BH and NS during the inspiral and post-merger phases \citep{2024arXiv240802534C}. During the inspiral phase, non-repeating FRB can be produced via black hole battery mechanism approximately $\sim1~\mathrm{ms} $ \citep{2015ApJ...814L..20M}, and escape from the system due to the low-medium density outside the system. During the post-merger stage, there are two scenarios. One is that the NS is plunged into BH without being tidally disrupted. If this is the case, only one FRB is produced when the magnetic field migrates from the NS to the BH until the NS is eventually absorbed, followed by the snapping of the magnetic field \citep{2015ApJ...814L..20M,2014A&A...562A.137F}. The other one is that NS is tidally disrupted by BH, where a high-density disk can be formed and a jet may be launched after merger. If this is the case, multiple FRBs can be produced via the synchrotron maser mechanism in the disk \citep{2019MNRAS.485.4091M, 2020MNRAS.494.4627M}, or ultra-relativistic magnetized shocks and magnetic reconnection in the jet \citep{2020ApJ...897....1L}.

\subsection{MCO as a massive neutron star scenario}
\textbf{(1). Inspiral phase:} the magnetospheres from the MCO (as a massive neutron star) and the neutron star will interact with each other when the distance between them becomes closer and closer due to the energy lost by gravitational waves and magnetic dipole radiation. Since the structure and interactions of magnetospheres from MCO and neutron star are very complicated, we only consider the dipole magnetic field within a vacuum rather than filled plasma \citep{2019MNRAS.483.2766L}. We can roughly think of a closed circuit between MCO and a neutron star, where the magnetic field lines are equivalent to loop wires, and the neutron star with its resistive crust as a battery which move toward each other with velocity $v=\beta c$ is immersed in the total magnetic field. If this is the case, the power of the circuit can be derived by applying Ohm's law:
\begin{equation}
    L_\mathrm{BNS}\sim \frac{\Delta\Phi^2}{\mathcal{R}_\mathrm{NS}}= \frac{c}{4\pi}\Delta\Phi^2,\label{LBNS}
\end{equation}
where $\Delta\Phi$ is emf drop due to magnetosphere interaction, and it can be expressed as
\begin{equation}
    \Delta\Phi\sim \frac{\Omega_{\rm orb}B_\mathrm{NS}B_\mathrm{MCO}}{cr\Big(\frac{B_\mathrm{NS}}{R_\mathrm{MCO}^3}+\frac{B_\mathrm{MCO}}{R_\mathrm{NS}^3}\Big)}.\label{BNSV}
\end{equation}
Here, $B_\mathrm{MCO}$ and $R_\mathrm{MCO}$ are the surface magnetic field strength and the radius of MCO, respectively. Based on the Eqs. (\ref{LBNS}) and (\ref{BNSV}), one has 
\begin{equation}
L_\mathrm{BNS}\sim\frac{G(M_\mathrm{NS}+M_\mathrm{MCO})}{4\pi cr^5}\frac{B_{\mathrm{NS}}^{2}B_{\mathrm{MCO}}^{2}}{\left(\frac{B_{\mathrm{NS}}}{R_{\mathrm{MCO}}^{3}}+\frac{B_{\mathrm{MCO}}}{R_{\mathrm{NS}}^{3}}\right)^{2}},\label{LBNS1}
\end{equation}
where $M_\mathrm{MCO}$ is the mass of MCO as a massive neutron star. By considering Eqs. (\ref{r}) and (\ref{LBNS1}), one can obtain $L_\mathrm{BNS}\propto (-t)^{-5/4}$ by adopting $M_{\rm MCO}$ to replace $M_{\rm BH}$ in Eq. (\ref{r}). Here, we ignore the uncertain effect of the magnetosphere, and the intrinsic luminosity may be higher than the luminosity predicted here or exhibit more complexity due to potential issues with magnetic reconnection \citep{2017JPlPh..83f6301L,2018JPlPh..84b6301L}. In order to roughly estimate the luminosity, we adopt some typical parameters of a neutron star (or massive neutron star) to perform the calculations, such as $M_\mathrm{MCO}=2.5M_\odot$, $C_\mathrm{MCO}=0.3$, $M_\mathrm{NS}=1.4M_\odot$, $R_\mathrm{NS}=10^6~\mathrm{cm}$ and $B_\mathrm{MCO}=B_\mathrm{NS}=10^{12}~\mathrm{G}$, one has luminosity $L_\mathrm{BNS}\sim1.05\times10^{43}~\mathrm{erg}~\mathrm{s}^{-1}$.

Similar to Section 3.1, by assuming pure curvature radiation as the main radiation mechanism of the precursor, one can plot the radiation spectrum of possible EM signals (e.g., $\gamma$-ray, X-ray, optical, and radio) which may be powered by magnetosphere interaction \citep{2021MNRAS.501.3184S, 2023PhRvL.130x5201M}. Fig \ref{fig:1} shows the predicted spectrum of multi-band EM radiations from the magnetosphere interaction mechanism by assuming a luminosity distance of 250 Mpc. We find that the radiated EM signals are too weak to be detected by Fermi/GBM, Fermi/LAT, Swift/BAT, Swift/XRT, Vera C. Rubin, and FAST by comparing the sensitivity of currently operating detectors. However, similar to the black hole battery scenario, only the high-energy (e.g., MeV-GeV) signals may be detected by Fermi/LAT.

\textbf{(2). Merger and post-merger phases:} another possible electromagnetic counterpart produced in the post-merger phase is kilonova emission. In order to calculate possible kilonova emission from the merger of a neutron star and MCO, one first needs to determine the remnant (e.g., black hole or massive NS) of such a merger system. In other words, one needs to compare the baryonic mass of the remnant ($M_\mathrm{b,rem}$) with the maximum baryonic mass of a neutron star at the mass-shedding limit ($M_\mathrm{b,max}$). We adopted an MCO mass of $2.2M_\odot$ (corresponding to $M_\mathrm{b}\sim 2.6 M_\odot$) without spin. A merger between this lower mass limit and a $1.4M_\odot$ (corresponding to $M_\mathrm{b}\sim 1.5 M_\odot$) neutron star would result in a remnant with $M_\mathrm{b,rem}\sim4.1M_\odot$, which exceeds the baryonic mass of a neutron star with a gravitational mass of $3M_\odot$ at the mass-shedding limit \citep{2020FrPhy..1524603G}. Although the mass of a hypermassive neutron star may exceed $3M_\odot$ depending on the uncertain equation of state (EoS), we still consider a prompt collapse into a black hole based on the above estimation. If this is the case, the disk mass can be calculated by the semi-analytical formula \citep{2019MNRAS.489L..91C}
\begin{equation}
\begin{aligned}
    \log_{10}\Bigg(\frac{M_{\rm disc}}{M_\odot}\Bigg)=&\\
    \max \Bigg(-3&,~b_1\Bigg(
    1 + b_2\tanh\Bigg[\frac{b_3-M_{\rm tot}/M_{\rm th}}{b_4}\Bigg]
    \Bigg)\Bigg).\label{NSdisk}
\end{aligned}
\end{equation}
Here, $M_{\rm tot}$ is the total mass of binary neutron stars, $M_{\rm th}$ is the threshold mass that separates the prompt collapse scenario from the delayed collapse scenario \citep{2013PhRvL.111m1101B}. In our calculations, we fixed $M_{\rm th}=3M_\odot$. $b_1=-31.335$, $b_2=-0.976$, $b_3=1.047$, and $b_4=0.059$ are the fitting parameters. For the semi-analytical formula of dynamical ejecta, we adopt the formula from \cite{2020PhRvD.101j3002K}, 
\begin{equation}
\begin{aligned}
    \frac{M_{\rm dyn}}{10^{-3}M_\odot}=M_1\Big(\frac{d_1}{C_1}+d_2\Big(\frac{M_2}{M_1}\Big)^n_0+d_3C_1\Big)+(1\leftrightarrow2).\label{NSdyn}
\end{aligned}
\end{equation}
$C_1$ is the compactness of $M_1$, and the best-fit parameters are $d_1=-9.334$, $d_2=114.17$, $d_3=-337.56$, and $n_0=-1.547$.

In some numerical simulations, a high $q$ which is defined as the mass ratio between MCO and neutron star shows that a light neutron star can be tidally disrupted by a more massive one. This behavior is similar to that of a black hole-neutron star merging system where tidal disruption has occurred \citep{2017PhRvD..95b4029D, 2020MNRAS.497.1488B}. However, the threshold value of $q$ depends on the EoS of a neutron star, and does not have a clear correlation with observational parameters. In that case, a threshold value of $q = 1.7$ is adopted based on the argument presented in  \cite{2020MNRAS.497.1488B}. It means that a binary system with mass ratio $q\geq1.7$ will lead to a TDE of the lighter one, followed by an accretion-induced prompt collapse of the massive one. Consequently, the ejecta mass will increase to $M_\mathrm{out} \sim 10^{-1}M_\odot$ which is comparable to that of TDE observed in black hole-neutron star mergers. By considering Eq. (\ref{NSdisk}) and (\ref{NSdyn}), one can roughly estimate the ejecta mass ($M_{\rm ej}=(0.4-4)\times10^{-4}~M_\odot$) of the merger of MCO and NS system whose mass ratio is below 1.7 \citep{2018ApJ...869..130R, 2020PhR...886....1N} by adopting $M_{\rm MCO}=2.5M_\odot$, $M_{\rm NS}=1.8M_\odot$, $R_{\rm NS}\approx 12.12~{\rm km}$, $C_{\rm MNS}=0.3$, and $f_{\rm wind}=0.04-0.4$ ($f_{\rm wind}=0$ corresponding to no ejecta). If this is the case, one can roughly calculate the kilonova emission shown in Figure \ref{fig:3} with $\beta=0.2$ and $\kappa=1~\text{g cm}^{-2}$ \citep{2021ApJ...912...14Y}.

\textbf{(3). Possible FRB emission:} a possible electromagnetic counterpart in radio band, FRB, may be produced before the merger in the binary system of MCO and neutron star. The mechanism of the emitted FRB arises from magnetosphere interaction which can produce multiple explosions, namely it possibly corresponds to the observed repeating FRBs \citep{2021MNRAS.501.3184S, 2023PhRvL.130x5201M}. In the post-merger stage, the no-repeating FRB may also be produced after the magnetic field from the two neutron stars that migrated to the newly formed black hole and finally escaped (Blitzar model).

\subsection{Identifying MCO in binary merging system of MCO and NS}
By summarizing Section 3.1 and Section 3.2, the properties of produced electromagnetic counterparts (e.g., precursor, kilonova, as well as possible FRB) from the merger of MCO (as a low-mass black hole) and NS may be different from that of binary system in MCO (as a massive NS) and NS. We will present details of the difference between those two types of binary systems.

(1) \emph{Precursor emission}: From the theoretical point of view, a possible precursor emission may be produced during the inspiral phase of the MCO (as a low-mass black hole or massive NS) merging with NS. Based on the black hole battery mechanism, the luminosity evolution can be expressed as $L_\mathrm{BH-NS}\sim(-t)^{-7/4}$ for the binary system of MCO as a low-mass black hole merging with NS. However, the luminosity evolution from the binary system of MCO as a massive NS merging with NS, $L_\mathrm{BNS}\sim(-t)^{-5/4}$, derived from the magnetosphere interactions, is different from that of binary system of MCO as a low-mass black hole merging with NS \citep{2023ApJ...956L..33M, 2021ApJ...921...92B, 2023ApJ...959...34B}. 

(2) \emph{Kilonova emission}: No matter whether MCO is a low-mass black hole or massive NS, kilonova emission can be produced in the post-merger phase of MCO and NS binary system. However, from the semi-analytic model of the numerical simulation, the ejecta mass of MCO as a low-mass black hole merging with NS is larger than that of a binary system with an MCO as a massive NS without TDE occurring. By considering the radioactive decay as the only source of kilonova energy \citep{2021ApJ...912...14Y}, the calculated kilonova emission from a low-mass black hole merging with the NS scenario is at least 6 times brighter than that of kilonova emission from a massive NS merging with a NS scenario. A similar result is reported by \cite{2021ApJ...911...87L} and \cite{2019ApJ...887L..35B}, who conclude that the peak luminosity and peak time of a Gap-NS system are different from those of an NS-NS system when the chirp mass falls within the range of $1.5-1.7~\rm M_\odot$.

(3) \emph{FRB}: For the binary system of MCO, a low-mass black hole and NS, one possible FRB can be produced during the inspiral phase via black hole battery mechanism while several FRBs can be produced at post-merger phase via 'Blitzar' mechanism and synchrotron maser mechanism. No matter which mechanism results in FRB emission, it should be a single explosion with no-repeating behavior. However, for the binary system of MCO a massive NS and NS, the repeating FRB-like signals can be produced from the interaction between those two magnetospheres before the merger, and another possible FRB can also be produced after the magnetic field from the two neutron stars that have migrated to the newly formed black hole. Unfortunately, such FRB emissions are very difficult to detect with current radio telescopes, even though it maybe probe to distinguish these two merger scenarios.

\section{MCO merger with a black hole} \label{sec:withBH}
In this section, we focus on discussing the possible electromagnetic signal emitted from the merger of MCO and a black hole with a mass range of $(5-30)~M_\odot$. There are also two types of objects of MCO, e.g., MCO as a low-mass black hole and a massive NS, respectively. Although some research has claimed that electromagnetic counterparts can be produced in binary black hole systems if one black hole is charged \citep{2016PhRvD..94f4046L,2016ApJ...827L..31Z,2017ApJ...839L...7D}. However, there is no evidence to support these arguments, because the LVK collaboration does not detect any electromagnetic counterparts from observed more than hundreds of observed binary black hole events \citep{2019PhRvX...9c1040A, 2021PhRvX..11b1053A, 2023PhRvX..13d1039A}. The absence of detected electromagnetic counterparts from the merger of binary black holes may be caused by two reasons, one is that the electromagnetic counterparts are too weak to be detected, and the other one is that no electromagnetic counterparts can be powered by such a merger of binary black holes. So, we do not discuss details about the case of MCO and the black hole system for MCO as a low-mass black hole. 

Alternatively, if the MCO is a massive neutron star merging with a black hole, electromagnetic counterparts may be produced in both the inspiral phase (caused by the black hole battery mechanism) and the post-merger phase (caused by tidal disruption). During the inspiral phase, the precursor emission of such a system is driven by the black hole battery mechanism. By adopting $a=0$ for black hole, and $B=10^{12}~\mathrm{G}$, $C_\mathrm{NS}=0.3$, $M_\mathrm{NS}=2.5M_\odot$ for massive neutron star, one can calculate the maximum luminosity which ranges from $L_\mathrm{BH-NS,max}\sim1.86\times10^{43}~\mathrm{erg/s}$ (for $M_\mathrm{BH}=5M_\odot$) to $L_\mathrm{BH-NS, max}\sim2.91\times10^{41}~\mathrm{erg/s}$ (for $M_\mathrm{BH}=30M_\odot$) at $r=R_\mathrm{sch}+R_\mathrm{NS}$ without TDE occurred. Here, $R_\mathrm{sch}$ is the Schwarzschild radius of a black hole. Moreover, one question is how strongly the adoption of different values of $a$ can affect the maximum luminosity. By adopting $a=0.99$, it is found that $L_\mathrm{BH-NS,max}$ is several times lower than that for $a=0$.

During the post-merger phase, it should be noted that the condition for TDE to occur becomes more stringent when the mass of the black hole and the compactness of the neutron star are increased. Figure \ref{fig:6} shows the lowest dimensionless spin of a black hole required for a TDE to occur which depends on black hole mass and compactness of the neutron star for a given $M_{\rm NS}=2.2M_\odot$, 2.5$M_\odot$, and 2.8$M_\odot$, and it requires a larger spin of black hole which can result in easier TDE occurred. If this is the case, kilonova emission may be powered when a TDE occurs for such a merger system. Based on the semi-analytical formula of Eqs. (\ref{out}) and (\ref{dyn}), one can roughly calculate the ejecta mass $M_\mathrm{ej}$ by adopting $M_\mathrm{NS}=2.5M_\odot$, $M_\mathrm{BH}=(5-30)M_\odot$, and $f_\mathrm{wind}=0.4$. For $a=0.99$, the calculated range of ejecta mass are $M_\mathrm{ej}=(0.09-0.25)M_\odot$ and $M_\mathrm{ej}=(0.05-0.27)M_\odot$ for $C_{\rm NS}=0.25$ and $C_{\rm NS}=0.3$, respectively. Also, if we change $a=0.8$, one can calculate the maximum ejecta mass $M_\mathrm{ej}=0.2M_\odot$ and $M_\mathrm{ej}=0.13M_\odot$ for $C_{\rm NS}=0.25$ and $C_{\rm NS}=0.3$, respectively. On the other hand, we adopt the semi-analytical formula to estimate the average speed of ejecta \citep{2016ApJ...825...52K}
\begin{equation}
    v_\mathrm{ave} = (0.01533/q + 0.1907) c\label{speed}.
\end{equation} 
Figure \ref{fig:7} shows the kilonova emission by adopting different values of $C_{\rm NS}=0.25$, 0.3 and $a=0.99$, 0.8 for $M_\mathrm{NS}=2.5\rm M_\odot$, $M_\mathrm{BH}=(5-30)\rm M_\odot$, and $f_\mathrm{wind}=0.4$. It is found that the kilonova emission is brighter than that of the situation in the merger system of MCO and NS due to a larger ejecta mass when a TDE occurs. Moreover, GW200115 and GW200105 did not observe possible kilonova emission because, without a TDE, a low dimensionless spin value resulted \citep{2021ApJ...915L...5A}. In our calculations, it requires a higher spin of the black hole to ensure that the TDE occurred.

\section{Conclusion and Discussion} \label{sec:end}
The first successful detection of GW emission resulting from the merger of binary black holes, known as the GW150914 event, confirmed general relativity and tested the theory of gravity \citep{2016PhRvL.116f1102A}. Coincident detections of EM and GW signals from the coalescences of neutron star binaries, known as the GW170817/GRB 170817A/AT2017gfo event, have the potential to provide an unparalleled understanding of r-process nucleosynthesis \citep{2017Natur.551...80K, 2017PASJ...69..102T, 2019Natur.574..497W}, Hubble’s constant \citep{2017PhRvL.119p1101A}, as well as the origin of short GRBs \citep{2017PhRvL.119p1101A, 2017ApJ...848L..14G, 2017ApJ...848L..15S, 2018NatCo...9..447Z}. Furthermore, several GW events from merger of NS-BH have been reported, but lack of any EM counterpart is detected because of low spin of black hole or significant localization error \citep{2021ApJ...915L...5A, 2024ApJ...970L..34A}. 

More interestingly, several compact objects in the binary system whose mass fall into the range of $(2-3)\rm~M_\odot$ have been determined through the GW emission \citep{2020ApJ...896L..44A}. However, whether it is a massive neutron star or a low-mass black hole, remain an open question and still under debate. In this paper, we propose a method to identify the object with the mass region of $(2.2-3)\rm~M_\odot$ in a binary system via the possible EM radiations before and after the mergers. One can obtain the following interesting results.
\begin{itemize}	
\item[$\bullet$] \textbf{MCO merges with a neutron star}: (1) A multi-band EM (called precursor) emission, including non-repeating FRB, can be produced during the inspiral phase due to the black hole battery mechanism when MCO is as a low-mass black hole, and the luminosity evolution of EM radiation can be expressed as $L\propto(-t)^{-7/4}$. On the other hand, during the post-merger phase, the bright kilonova emission which is powered by radioactive decay of ejecta, and non-repeating FRB caused by 'Blitzar' mechanism and synchrotron maser mechanism, can be expected to be detected by ground-based telescopes. (2) A multi-band precursor emission, including repeating FRB, can also be produced during the inspiral phase due to the magnetospheres interaction mechanism when MCO is as a massive neutron star, and the luminosity evolution of EM radiation can be expressed as $L\propto(-t)^{-5/4}$. Moreover, a dark or bright kilonova emission can be powered during the post-merger phase when the mass ratio $q$ is below or above 1.7, respectively. In any case, no matter which mechanism we adopt, such precursor emissions are too weak to be detected by most current telescopes.
\item[$\bullet$] \textbf{MCO merges with a black hole:} By considering the merger system between MCO and a black hole when MCO is a massive NS, we find that it requires the black hole with high spin (e.g., $a\sim0.8-0.99$) to make sure the TDE occurred. The luminosity evolution of precursor emission can also be expressed as $L\propto(-t)^{-7/4}$ which decreases with the increasing mass of the black hole during the inspiral phase. A bright kilonova emission caused by a larger ejecta mass may be powered during the post-merger phase.
\end{itemize}

The GW events of GW190814 and GW230529 are the MCO merging with a black hole or a neutron star, respectively, and no associated electromagnetic counterparts were observed for those two cases. However, one can adopt our method to identify the MCO in the GW190814-like and GW230529-like events in the future if the electromagnetic counterparts are luckily to be observed during the inspiral, merger and ringdown phases of these binaries. For example, if one can observe the electromagnetic counterparts of GW190814-like event, the MCO should be a neutron star in this binary system. That is because no observational evidence to indicate that a BH-BH merger can produce the observed EM signals. For GW230529-like event, if one can observe the repeating FRB and $L\propto(-t)^{-5/4}$ during inspiral phase, and a dark kilonova emission during post-merger phase, the MCO should be a massive neutron star. On the contrary, if one observe the non-repeating FRB and $L\propto(-t)^{-7/4}$ during inspiral phase, and a bright kilonova emission during post-merger phase, the MCO should be a low-mass BH. The summary of possible EM radiations and their characteristics are listed in Table \ref{table1}.

It is worth noting that the luminosity evolution of precursor emissions during the inspiral phase is calculated by adopting some approximations. The intrinsic luminosity of precursor emissions for the black hole battery mechanism may be dimmer than that calculated here due to the neglected magnetosphere interaction of the neutron star. However, it may be underestimated by adopting vacuum approximation in binary neutron star systems. On the other hand, we only consider the black hole battery mechanism and magnetosphere interaction mechanism to do the calculations in this paper, and there may exist other mechanisms that can result in different precursor emissions \citep{2012PhRvL.108a1102T, 2022MNRAS.514.5385N}. For estimated kilonova emission, we adopt a semi-analytic model to calculate the ejecta mass of BH-NS and NS-NS systems, and we do not consider the effects from density distribution, spatial distribution, and velocity distribution of ejecta mass, as well as the effects from EOS of NS and different sources of kilonova emission. So, the intrinsic kilonova emission may exist a slight difference from what we calculated, but it needs to be confirmed through numerical simulations of magnetohydrodynamics. Furthermore, even the FRB associated with any binary systems is not detected, it is a good probe to identify the binary systems if such an associated event is lucky to be detected in the future. 

Moreover, if one can identify the MCO with the mass region of (2.2-3)$M_\odot$ as a massive NS, it is very important to constrain the poor understanding EOS of NS \citep{2020PhRvL.125z1104T,2024ApJ...962...61M}. Alternatively, if the MCO is identified as a low-mass BH with the mass region of (2.2-3)$M_\odot$, it is very helpful to understand the formation of such a low-mass black hole, e.g., primordial black hole \citep{2016PhRvL.117f1101S, 2023JCAP...11..069J}, or remnant of NS-NS merger \citep{2017ApJ...844L..19P, 2024ApJ...971L..24D, 2024PhRvD.109l3011S}. 

Besides adopting EM signals to distinguish the identity of MCO, previous studies adopted the GW data \citep{2012PhRvD..85l3007D, 2015ApJ...807L..24L, 2018ApJ...856..110Y,2020ApJ...904...80E,2020PhRvD.102b3025F,2020ApJ...893L..41C,2022ApJ...931..108F, 2022ApJ...941...98B, 2022PhRvD.105f4063C} or the tidal disruption \citep{2015PhRvD..92h1504P, 2023ApJ...949L...6C} to distinguish the nature of compact star. By comparing the EM method with other GW method, the advantage of GW method is to confirm the mass of binary star, but the disadvantage is that it is difficult to identify the type of star if the star mass is in the range of (2.2-3)$M_\odot$. In order to confirm the type of star in such mass range, one has to measure the higher order terms of GW emission which are unlikely to be detected by the current GW detectors (e.g., LIGO and Virgo), but which could be captured by the third-generation GW detectors, such as Cosmic Explorer and Einstein Telescope. For EM method, it can help us distinguish the MCO within the mass range of (2.2-3)$M_\odot$ as the NS or BH when the precursor, FRB, as well as kilonova can be simultaneously detected. However, the disadvantage is the poorly understood equation of state of NS and dependence on model parameters. Therefore, the optimal strategy should be to detect both EM and GW emissions. 

From an observational point of view, it is very difficult to simultaneously detect the precursor, FRB, and kilonova. For example, the precursor emission may only be detectable by Fermi/LAT, as it is too weak to be observed by Fermi/GBM, Swift/BAT, Swift/XRT, the Vera C. Rubin, or FAST. An expectation is that the GW event is bright enough and nearby, and also requires those telescopes to observed the same location of GW event before the GW event happened. The FRB may be detected by FAST, but it require the observations before the GW event occurs. For kilonova emission, one of the main scientific goals of the space-based multi-band Astronomical Variable Objects Monitor (SVOM), is to conduct follow-up observations of kilonova candidates via its visible-band telescope (VT) together with other optical survey projects after the merger of binary star \citep{2016arXiv161006892W}. Thus, it is expected that SVOM will be helpful in making deep observations of kilonova candidates in the future.

\begin{acknowledgments}
We thank Qiuhong Chen and Tian-Ci Liu for helpful discussions. This work is supported by the Guangxi Science Foundation the National (grant No. 2023GXNSFDA026007), the Natural Science Foundation of China (grant Nos. 12494574, 11922301 and 12133003), the Program of Bagui Scholars Program (LHJ), and the Guangxi Talent Program (“Highland of Innovation Talents”).
\end{acknowledgments}

\newpage
\begin{figure}
\center
\includegraphics[scale = 0.99]{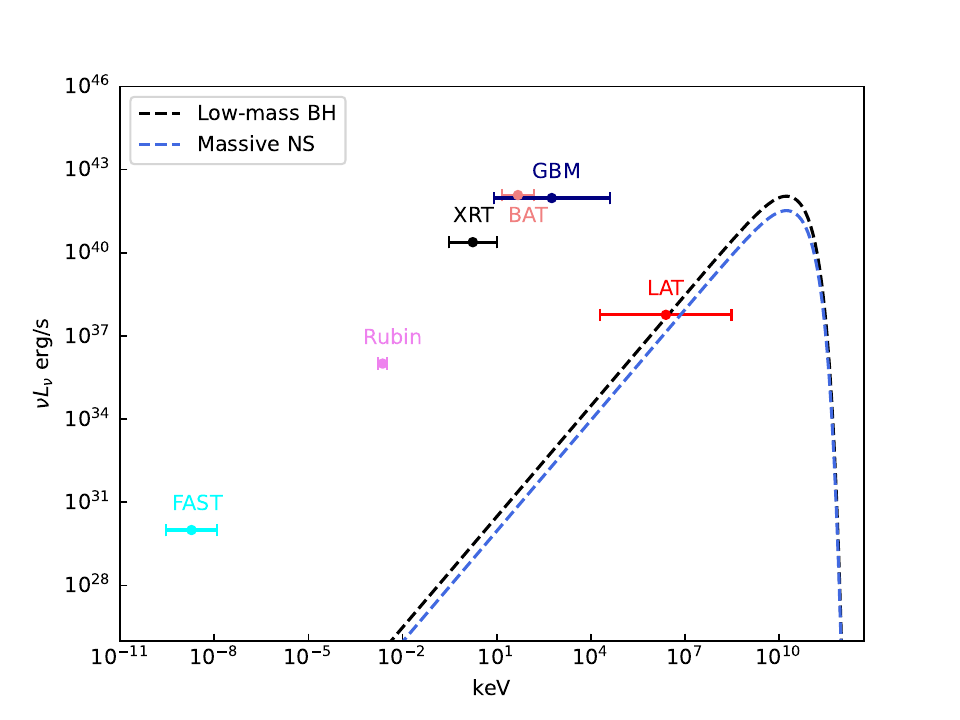}
\hfill\caption{The spectrum of predicted precursor emissions from the merger of MCO and NS during the inspiral phase by adopting the parameters of NS $M_{\rm NS}=1.4M_\odot$, $R_{\rm NS}=10^6~{\rm cm}$, and $B=10^{12}~{\rm G}$. Blue and black dashed lines are corresponding to MCO as a massive NS ($C_{\rm NS}=0.3$ and $B_{\rm MCO}=10^{12}~{\rm G}$) and a low-mass BH ($M_{\rm MCO}=2.5M_\odot$ and $a=0$), respectively. The separation radius is adopted as $6GM_{\rm MCO}/c^2\sim 2.2\times10^6~{\rm cm}$. The color solid circles are the sensitivity of each detector.}
\label{fig:1}
\end{figure}
\begin{figure}
\center
\includegraphics[scale = 0.5]{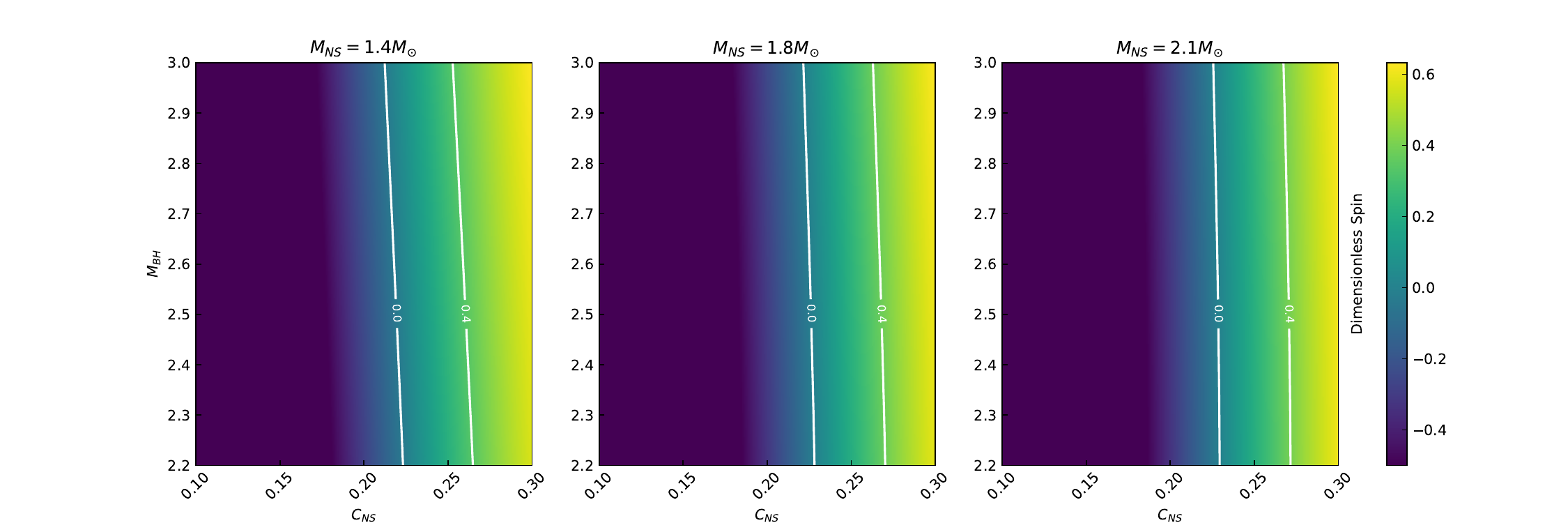}
\hfill\caption{Two-dimensional distributions of mass of BH and compactness of NS for given the mass of NS ($M_{\rm NS}=1.4M_\odot$, $1.8M_\odot$, and $2.1M_\odot$). The two white contour lines represent the thresholds for TDE occurred at $a=0$ and $a=0.4$, respectively.}
\label{fig:2}
\end{figure}
\begin{figure}
\center
\includegraphics[scale = 0.8]{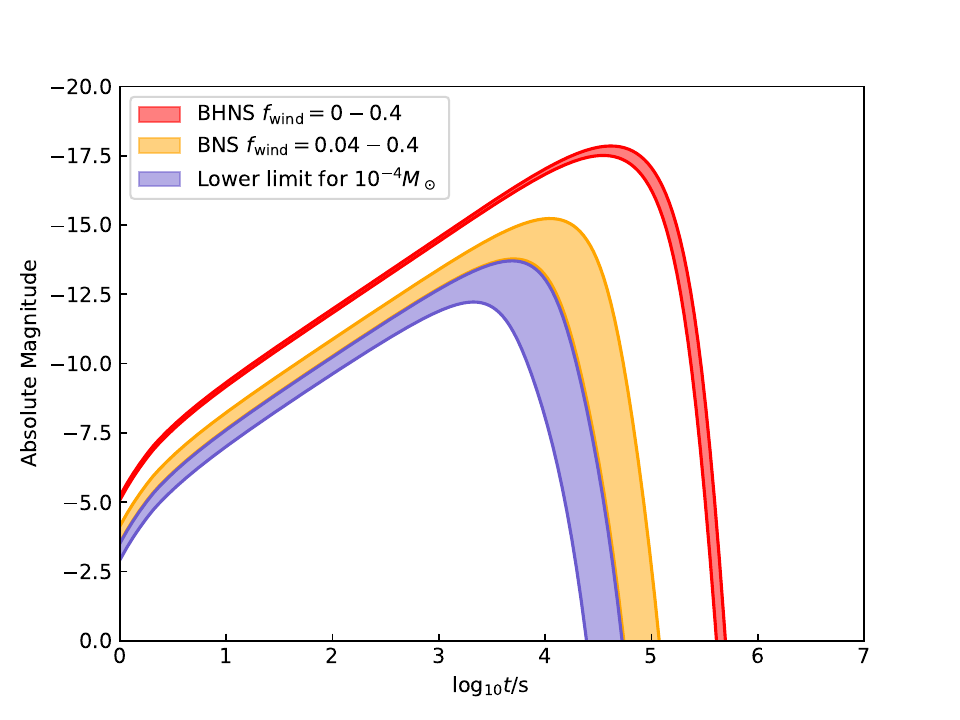}
\caption{Kilonova emission in g-band from merger of MCO-NS. Red region is the MCO as a low-mass BH with $f_\mathrm{wind}=(0-0.4)$. Orange region is the MCO as a massive NS with $f_\mathrm{wind}=(0.04-0.4)$. Purple region is $M_{\rm disk}=10^{-4}M_\odot$ as the lower limit of ejecta mass in prompt collapse case when MCO is as a massive NS. In each case, we adopt the same mass of MCO as $2.5\rm M_\odot$ and neutron star as $1.8\rm M_\odot$. Dimensionless spin of low mass black hole is $0.4$. Compactness of massive neutron star is $0.3$. Raidius of neutron star is $12.12\rm ~km$.}
\label{fig:3}
\end{figure}
\begin{figure}
\center
\includegraphics[scale = 0.5]{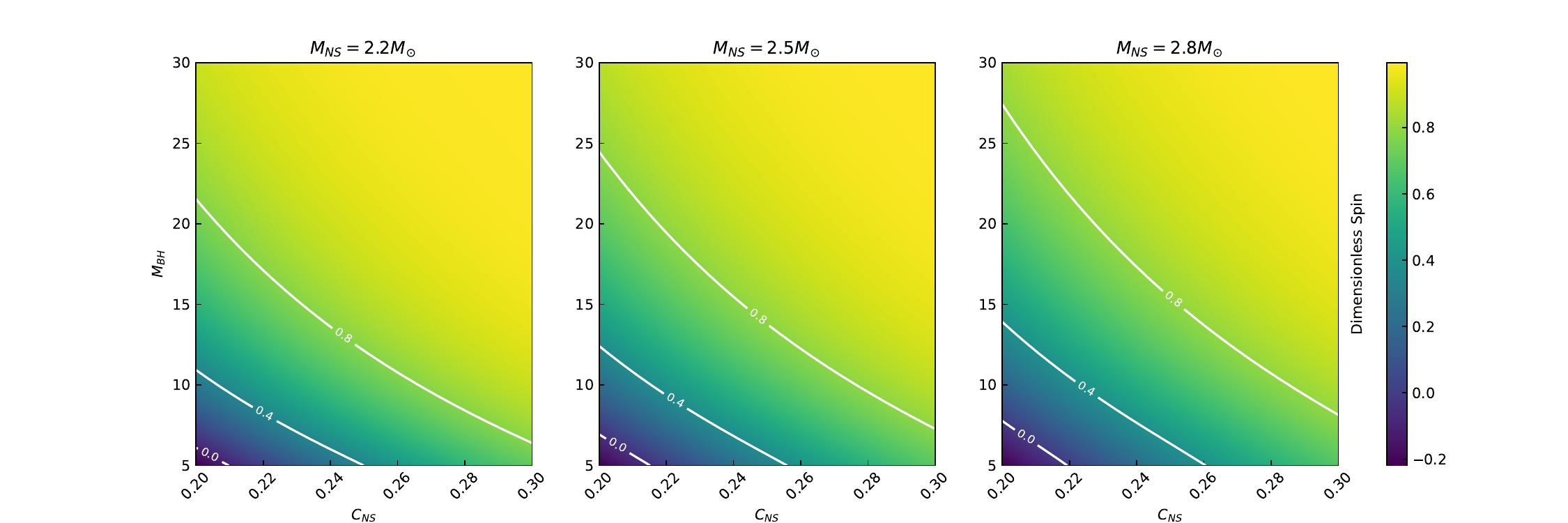}
\hfill\caption{Similar to Fig \ref{fig:2}, but adopting $M_{\rm BH}=(5-30)M_\odot$ and $C_{\rm NS}=(0.2-0.3)$. The three white contour lines represent the thresholds for TDE occurred at at $a=0$, $a=0.4$, and $a=0.8$, respectively.}
\label{fig:6}
\end{figure}
\begin{figure}
\center
\includegraphics[scale = 0.5]{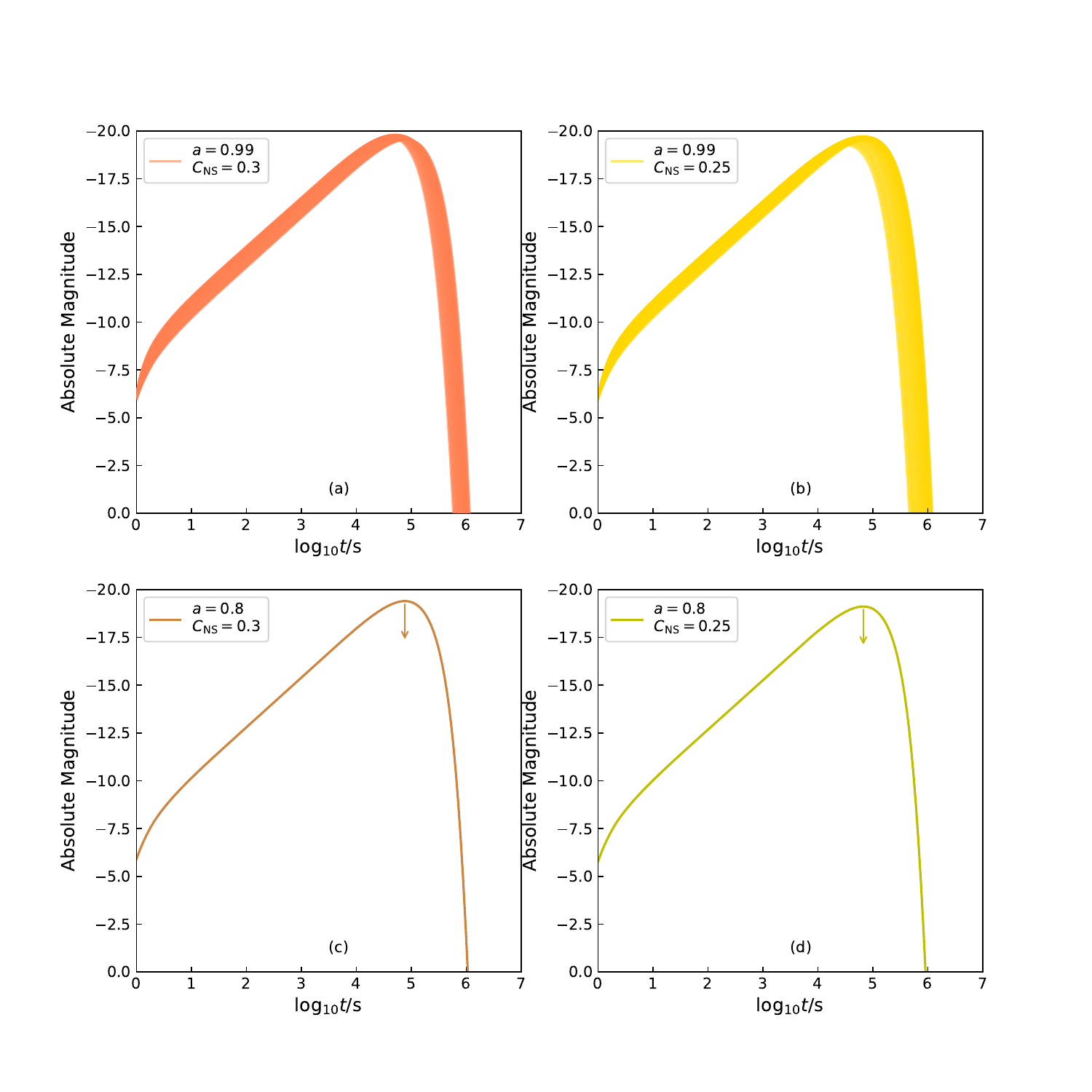}
\hfill\caption{Kilonova emissions in g-band from merger of MCO-BH by adopting $\kappa=1~\text{g cm}^{-2}$, $M_\mathrm{MCO}=2.5M_\odot$, $M_\mathrm{BH}=(5-30)M_\odot$, and different values of $a$ and $C_{\rm NS}$. (a) $a=0.99$ and $C_{\rm NS}=0.3$; (b) $a=0.99$ and $C_{\rm NS}=0.25$; (c) $a=0.8$ and $C_{\rm NS}=0.3$; (d) $a=0.8$ and $C_{\rm NS}=0.25$. In figure (c) and (d), Only maximum kilonova emission is plotted due to not all BH mass parameters resulting in a non-zero $M_\mathrm{ej}$.}
\label{fig:7}
\end{figure}
\newpage
\begin{table*}[h]\footnotesize 
  \centering
  \caption{Summarized to the possible EM counterparts and its characteristic from merger of MCO and BH (or NS).}
  \renewcommand{\arraystretch}{1.5}
  \setlength{\tabcolsep}{4.0mm}{
  \begin{center}
  \begin{tabular}{|c|c|P{5cm}|P{4cm}|}
  \hline
  Companion Object & MCO & EM Counterparts/mechanism & Characteristic \\
  \hline
  \multirow{6}{*}{NS} & \multirow{3}{*}{low-mass BH} & \emph{inspiral}: multi-band precursor/BH battery mechanism & $L\propto(-t)^{-7/4}$ \\
  \cline{3-4}
   & & \emph{post-merger}: kilonova/radioactive decay & bright for all $q$\\
   \cline{3-4}
   & & FRB: BH battery mechanism/Blitzar mechanism & Non-repeating in both inspiral and post-merger\\
   \cline{2-4}
   & \multirow{3}{*}{massive NS} & \emph{inspiral}: multi-band precursor/magnetosphere interaction & $L\propto(-t)^{-5/4}$\\
   \cline{3-4}
   & & \emph{post-merger}: kilonova/radioactive decay & dark for $q<1.7$, bright for $q>1.7$\\
   \cline{3-4}
   & & FRB/magnetosphere interaction mechanism/Blitzar mechanism & repeating during inspiral, and Non-repeating during post-merger\\
   \hline
   \multirow{5}{*}{BH} & \multirow{2}{*}{low-mass BH} & \multirow{2}{*}{No EM counterparts} & \multirow{2}{*}{No discussion}\\
    & & & \\
   \cline{2-4}
    & \multirow{3}{*}{massive NS} & \emph{inspiral}: multi-band precursor/BH battery mechanism & $L\propto(-t)^{-7/4}$, the dimmer, the more massive BH \\
    \cline{3-4}
    & & \emph{post-merger}: kilonova/radioactive decay & bright for limit of high $a$ \\
    \cline{3-4}
    & & FRB: BH battery mechanism/Blitzar mechanism & No-repeating in both inspiral and post-merger\\
    \hline
\end{tabular}\label{table1}
\end{center}}
\end{table*}


\begin{thebibliography}{99}
\bibitem[Abac et al.(2024)]{2024ApJ...970L..34A} Abac, A.~G., Abbott, R., Abouelfettouh, I., et al.\ 2024, \apjl, 970, L34. doi:10.3847/2041-8213/ad5beb
\bibitem[Abbott et al.(2019)]{2019PhRvX...9c1040A} Abbott, B.~P., Abbott, R., Abbott, T.~D., et al.\ 2019, Physical Review X, 9, 031040. doi:10.1103/PhysRevX.9.031040
\bibitem[Abbott et al.(2017a)]{2017PhRvL.119p1101A} Abbott, B.~P., Abbott, R., Abbott, T.~D., et al.\ 2017a, \prl, 119, 161101. doi:10.1103/PhysRevLett.119.161101
\bibitem[Abbott et al.(2017b)]{2017ApJ...848L..12A} Abbott, B.~P., Abbott, R., Abbott, T.~D., et al.\ 2017b, \apjl, 848, L12. doi:10.3847/2041-8213/aa91c9
\bibitem[Abbott et al.(2016)]{2016PhRvL.116f1102A} Abbott, B.~P., Abbott, R., Abbott, T.~D., et al.\ 2016, \prl, 116, 061102. doi:10.1103/PhysRevLett.116.061102
\bibitem[Abbott et al.(2020a)]{2020ApJ...896L..44A} Abbott, R., Abbott, T.~D., Abraham, S., et al.\ 2020a, \apjl, 896, L44. doi:10.3847/2041-8213/ab960f
\bibitem[Abbott et al.(2020b)]{2020ApJ...892L...3A} Abbott, B.~P., Abbott, R., Abbott, T.~D., et al.\ 2020b, \apjl, 892, L3. doi:10.3847/2041-8213/ab75f5
\bibitem[Abbott et al.(2021a)]{2021PhRvX..11b1053A} Abbott, R., Abbott, T.~D., Abraham, S., et al.\ 2021a, Physical Review X, 11, 021053. doi:10.1103/PhysRevX.11.021053
\bibitem[Abbott et al.(2021b)]{2021ApJ...915L...5A} Abbott, R., Abbott, T.~D., Abraham, S., et al.\ 2021b, \apjl, 915, L5. doi:10.3847/2041-8213/ac082e
\bibitem[Abbott et al.(2023)]{2023PhRvX..13d1039A} Abbott, R., Abbott, T.~D., Acernese, F., et al.\ 2023, Physical Review X, 13, 041039. doi:10.1103/PhysRevX.13.041039
\bibitem[Ai et al.(2020)]{2020ApJ...893..146A} Ai, S., Gao, H., \& Zhang, B.\ 2020, \apj, 893, 146. doi:10.3847/1538-4357/ab80bd
\bibitem[Alexander et al.(2021)]{2021ApJ...923...66A} Alexander, K.~D., Schroeder, G., Paterson, K., et al.\ 2021, \apj, 923, 66. doi:10.3847/1538-4357/ac281a
\bibitem[Anand et al.(2021)]{2021NatAs...5...46A} Anand, S., Coughlin, M.~W., Kasliwal, M.~M., et al.\ 2021, Nature Astronomy, 5, 46. doi:10.1038/s41550-020-1183-3
\bibitem[Barbieri et al.(2019a)]{2019ApJ...887L..35B} Barbieri, C., Salafia, O.~S., Colpi, M., et al.\ 2019a, \apjl, 887, L35. doi:10.3847/2041-8213/ab5c1e
\bibitem[Barbieri et al.(2019b)]{2019A&A...625A.152B} Barbieri, C., Salafia, O.~S., Perego, A., et al.\ 2019b, \aap, 625, A152. doi:10.1051/0004-6361/201935443
\bibitem[Bardeen et al.(1972)]{1972ApJ...178..347B} Bardeen, J.~M., Press, W.~H., \& Teukolsky, S.~A.\ 1972, \apj, 178, 347. doi:10.1086/151796
\bibitem[Bauswein et al.(2013)]{2013PhRvL.111m1101B} Bauswein, A., Baumgarte, T.~W., \& Janka, H.-T.\ 2013, \prl, 111, 131101. doi:10.1103/PhysRevLett.111.131101
\bibitem[Bauswein et al.(2017)]{2017ApJ...850L..34B} Bauswein, A., Just, O., Janka, H.-T., et al.\ 2017, \apjl, 850, L34. doi:10.3847/2041-8213/aa9994
\bibitem[Beloborodov(2021)]{2021ApJ...921...92B} Beloborodov, A.~M.\ 2021, \apj, 921, 92. doi:10.3847/1538-4357/ac17e7
\bibitem[Beloborodov(2023)]{2023ApJ...959...34B} Beloborodov, A.~M.\ 2023, \apj, 959, 34. doi:10.3847/1538-4357/acf659
\bibitem[Berger(2014)]{2014ARA&A..52...43B} Berger, E.\ 2014, \araa, 52, 43. doi:10.1146/annurev-astro-081913-035926
\bibitem[Bernuzzi et al.(2020)]{2020MNRAS.497.1488B} Bernuzzi, S., Breschi, M., Daszuta, B., et al.\ 2020, \mnras, 497, 1488. doi:10.1093/mnras/staa1860
\bibitem[Biswas et al.(2021)]{2021MNRAS.505.1600B} Biswas, B., Nandi, R., Char, P., et al.\ 2021, \mnras, 505, 1600. doi:10.1093/mnras/stab1383
\bibitem[Breu \& Rezzolla(2016)]{2016MNRAS.459..646B} Breu, C. \& Rezzolla, L.\ 2016, \mnras, 459, 646. doi:10.1093/mnras/stw575
\bibitem[Broderick et al.(2000)]{2000ApJ...537..351B} Broderick, A., Prakash, M., \& Lattimer, J.~M.\ 2000, \apj, 537, 351. doi:10.1086/309010
\bibitem[Brown et al.(2022)]{2022ApJ...941...98B} Brown, S.~M., Capano, C.~D., \& Krishnan, B.\ 2022, \apj, 941, 98. doi:10.3847/1538-4357/ac98fe
\bibitem[Cardall et al.(2001)]{2001ApJ...554..322C} Cardall, C.~Y., Prakash, M., \& Lattimer, J.~M.\ 2001, \apj, 554, 322. doi:10.1086/321370
\bibitem[Carrasco et al.(2021)]{2021PhRvD.104f3004C} Carrasco, F., Shibata, M., \& Reula, O.\ 2021, \prd, 104, 063004. doi:10.1103/PhysRevD.104.063004
\bibitem[Chen \& Chatziioannou(2020)]{2020ApJ...893L..41C} Chen, H.-Y. \& Chatziioannou, K.\ 2020, \apjl, 893, L41. doi:10.3847/2041-8213/ab86bc
\bibitem[Chen et al.(2024)]{2024MNRAS.529.1154C} Chen, M.-H., Li, L.-X., Chen, Q.-H., et al.\ 2024, \mnras, 529, 1154. doi:10.1093/mnras/stae475
\bibitem[Clarke et al.(2023)]{2023ApJ...949L...6C} Clarke, T.~A., Chastain, L., Lasky, P.~D., et al.\ 2023, \apjl, 949, L6. doi:10.3847/2041-8213/acd33b
\bibitem[Clarke et al.(2024)]{2024arXiv240802534C} Clarke, T.~A., Sarin, N., Howell, E.~J., et al.\ 2024, arXiv:2408.02534. doi:10.48550/arXiv.2408.02534
\bibitem[Connaughton et al.(2016)]{2016ApJ...826L...6C} Connaughton, V., Burns, E., Goldstein, A., et al.\ 2016, \apjl, 826, L6. doi:10.3847/2041-8205/826/1/L6
\bibitem[Cook et al.(1994)]{1994ApJ...424..823C} Cook, G.~B., Shapiro, S.~L., \& Teukolsky, S.~A.\ 1994, \apj, 424, 823. doi:10.1086/173934
\bibitem[Coughlin et al.(2019)]{2019MNRAS.489L..91C} Coughlin, M.~W., Dietrich, T., Margalit, B., et al.\ 2019, \mnras, 489, L91. doi:10.1093/mnrasl/slz133
\bibitem[Coupechoux et al.(2022)]{2022PhRvD.105f4063C} Coupechoux, J.-F., Arbey, A., Chierici, R., et al.\ 2022, \prd, 105, 064063. doi:10.1103/PhysRevD.105.064063
\bibitem[D'Orazio et al.(2016)]{2016PhRvD..94b3001D} D'Orazio, D.~J., Levin, J., Murray, N.~W., et al.\ 2016, \prd, 94, 023001. doi:10.1103/PhysRevD.94.023001
\bibitem[D'Orazio et al.(2022)]{2022ApJ...927...56D} D'Orazio, D.~J., Haiman, Z., Levin, J., et al.\ 2022, \apj, 927, 56. doi:10.3847/1538-4357/ac4bdb
\bibitem[Damour et al.(2012)]{2012PhRvD..85l3007D} Damour, T., Nagar, A., \& Villain, L.\ 2012, \prd, 85, 123007. doi:10.1103/PhysRevD.85.123007
\bibitem[de Mink \& King(2017)]{2017ApJ...839L...7D} de Mink, S.~E. \& King, A.\ 2017, \apjl, 839, L7. doi:10.3847/2041-8213/aa67f3
\bibitem[Deb et al.(2021)]{2021ApJ...922..149D} Deb, D., Mukhopadhyay, B., \& Weber, F.\ 2021, \apj, 922, 149. doi:10.3847/1538-4357/ac222a
\bibitem[Dietrich et al.(2017)]{2017PhRvD..95b4029D} Dietrich, T., Ujevic, M., Tichy, W., et al.\ 2017, \prd, 95, 024029. doi:10.1103/PhysRevD.95.024029
\bibitem[Drout et al.(2017)]{2017Sci...358.1570D} Drout, M.~R., Piro, A.~L., Shappee, B.~J., et al.\ 2017, Science, 358, 1570. doi:10.1126/science.aaq0049
\bibitem[DuPont \& MacFadyen(2024)]{2024ApJ...971L..24D} DuPont, M. \& MacFadyen, A.\ 2024, \apjl, 971, L24. doi:10.3847/2041-8213/ad66d2
\bibitem[D{\'\i}az et al.(2017)]{2017ApJ...848L..29D} D{\'\i}az, M.~C., Macri, L.~M., Garcia Lambas, D., et al.\ 2017, \apjl, 848, L29. doi:10.3847/2041-8213/aa9060
\bibitem[Essick \& Landry(2020)]{2020ApJ...904...80E} Essick, R. \& Landry, P.\ 2020, \apj, 904, 80. doi:10.3847/1538-4357/abbd3b
\bibitem[Falcke \& Rezzolla(2014)]{2014A&A...562A.137F} Falcke, H. \& Rezzolla, L.\ 2014, \aap, 562, A137. doi:10.1051/0004-6361/201321996
\bibitem[Farah et al.(2022)]{2022ApJ...931..108F} Farah, A., Fishbach, M., Essick, R., et al.\ 2022, \apj, 931, 108. doi:10.3847/1538-4357/ac5f03
\bibitem[Fasano et al.(2020)]{2020PhRvD.102b3025F} Fasano, M., Wong, K.~W.~K., Maselli, A., et al.\ 2020, \prd, 102, 023025. doi:10.1103/PhysRevD.102.023025
\bibitem[Fern{\'a}ndez et al.(2017)]{2017CQGra..34o4001F} Fern{\'a}ndez, R., Foucart, F., Kasen, D., et al.\ 2017, Classical and Quantum Gravity, 34, 154001. doi:10.1088/1361-6382/aa7a77
\bibitem[Fishbach et al.(2019)]{2019ApJ...871L..13F} Fishbach, M., Gray, R., Maga{\~n}a Hernandez, I., et al.\ 2019, \apjl, 871, L13. doi:10.3847/2041-8213/aaf96e
\bibitem[Foucart et al.(2018)]{2018PhRvD..98h1501F} Foucart, F., Hinderer, T., \& Nissanke, S.\ 2018, \prd, 98, 081501. doi:10.1103/PhysRevD.98.081501
\bibitem[Gao et al.(2020)]{2020FrPhy..1524603G} Gao, H., Ai, S.-K., Cao, Z.-J., et al.\ 2020, Frontiers of Physics, 15, 24603. doi:10.1007/s11467-019-0945-9
\bibitem[Gao et al.(2017)]{2017ApJ...837...50G} Gao, H., Zhang, B., L{\"u}, H.-J., et al.\ 2017, \apj, 837, 50. doi:10.3847/1538-4357/aa5be3
\bibitem[Goldstein et al.(2017)]{2017ApJ...848L..14G} Goldstein, A., Veres, P., Burns, E., et al.\ 2017, \apjl, 848, L14. doi:10.3847/2041-8213/aa8f41
\bibitem[Gompertz et al.(2020)]{2020MNRAS.497..726G} Gompertz, B.~P., Cutter, R., Steeghs, D., et al.\ 2020, \mnras, 497, 726. doi:10.1093/mnras/staa1845
\bibitem[Hotokezaka et al.(2013)]{2013PhRvD..87b4001H} Hotokezaka, K., Kiuchi, K., Kyutoku, K., et al.\ 2013, \prd, 87, 024001. doi:10.1103/PhysRevD.87.024001
\bibitem[Howlett \& Davis(2020)]{2020MNRAS.492.3803H} Howlett, C. \& Davis, T.~M.\ 2020, \mnras, 492, 3803. doi:10.1093/mnras/staa049
\bibitem[Hu et al.(2017)]{2017SciBu..62.1433H} Hu, L., Wu, X., Andreoni, I., et al.\ 2017, Science Bulletin, 62, 1433. doi:10.1016/j.scib.2017.10.006
\bibitem[Jangra et al.(2023)]{2023JCAP...11..069J} Jangra, P., Kavanagh, B.~J., \& Diego, J.~M.\ 2023, \jcap, 2023, 069. doi:10.1088/1475-7516/2023/11/069
\bibitem[Jin et al.(2018)]{2018ApJ...857..128J} Jin, Z.-P., Li, X., Wang, H., et al.\ 2018, \apj, 857, 128. doi:10.3847/1538-4357/aab76d
\bibitem[Jin et al.(2016)]{2016NatCo...712898J} Jin, Z.-P., Hotokezaka, K., Li, X., et al.\ 2016, Nature Communications, 7, 12898. doi:10.1038/ncomms12898
\bibitem[Kalogera \& Baym(1996)]{1996ApJ...470L..61K} Kalogera, V. \& Baym, G.\ 1996, \apjl, 470, L61. doi:10.1086/310296
\bibitem[Kasen et al.(2017)]{2017Natur.551...80K} Kasen, D., Metzger, B., Barnes, J., et al.\ 2017, \nat, 551, 80. doi:10.1038/nature24453
\bibitem[Kawaguchi et al.(2016)]{2016ApJ...825...52K} Kawaguchi, K., Kyutoku, K., Shibata, M., et al.\ 2016, \apj, 825, 52. doi:10.3847/0004-637X/825/1/52
\bibitem[Kawaguchi et al.(2020)]{2020ApJ...889..171K} Kawaguchi, K., Shibata, M., \& Tanaka, M.\ 2020, \apj, 889, 171. doi:10.3847/1538-4357/ab61f6
\bibitem[Khosravi Largani et al.(2022)]{2022MNRAS.515.3539K} Khosravi Largani, N., Fischer, T., Sedrakian, A., et al.\ 2022, \mnras, 515, 3539. doi:10.1093/mnras/stac1916
\bibitem[Kr{\"u}ger \& Foucart(2020)]{2020PhRvD.101j3002K} Kr{\"u}ger, C.~J. \& Foucart, F.\ 2020, \prd, 101, 103002. doi:10.1103/PhysRevD.101.103002
\bibitem[Lai(2012)]{2012ApJ...757L...3L} Lai, D.\ 2012, \apjl, 757, L3. doi:10.1088/2041-8205/757/1/L3
\bibitem[Lattimer \& Prakash(2001)]{2001ApJ...550..426L} Lattimer, J.~M. \& Prakash, M.\ 2001, \apj, 550, 426. doi:10.1086/319702
\bibitem[Li \& Paczy{\'n}ski(1998)]{1998ApJ...507L..59L} Li, L.-X. \& Paczy{\'n}ski, B.\ 1998, \apjl, 507, L59. doi:10.1086/311680
\bibitem[Li \& Shen(2021)]{2021ApJ...911...87L} Li, Y. \& Shen, R.-F.\ 2021, \apj, 911, 87. doi:10.3847/1538-4357/abe462
\bibitem[Liebling \& Palenzuela(2016)]{2016PhRvD..94f4046L} Liebling, S.~L. \& Palenzuela, C.\ 2016, \prd, 94, 064046. 
doi:10.1103/PhysRevD.94.064046
\bibitem[Littenberg et al.(2015)]{2015ApJ...807L..24L} Littenberg, T.~B., Farr, B., Coughlin, S., et al.\ 2015, \apjl, 807, L24. doi:10.1088/2041-8205/807/2/L24
\bibitem[Lopes \& Menezes(2022)]{2022ApJ...936...41L} Lopes, L.~L. \& Menezes, D.~P.\ 2022, \apj, 936, 41. doi:10.3847/1538-4357/ac81c4
\bibitem[Lyubarsky(2020)]{2020ApJ...897....1L} Lyubarsky, Y.\ 2020, \apj, 897, 1. doi:10.3847/1538-4357/ab97b5
\bibitem[Lyutikov et al.(2018)]{2018JPlPh..84b6301L} Lyutikov, M., Komissarov, S., Sironi, L., et al.\ 2018, Journal of Plasma Physics, 84, 635840201. doi:10.1017/S0022377818000168
\bibitem[Lyutikov et al.(2017)]{2017JPlPh..83f6301L} Lyutikov, M., Sironi, L., Komissarov, S.~S., et al.\ 2017, Journal of Plasma Physics, 83, 635830601. doi:10.1017/S0022377817000629
\bibitem[Lyutikov(2019)]{2019MNRAS.483.2766L} Lyutikov, M.\ 2019, \mnras, 483, 2766. doi:10.1093/mnras/sty3303
\bibitem[L{\"u} et al.(2019)]{2019MNRAS.486.4479L} L{\"u}, H.-J., Shen, J., Lan, L., et al.\ 2019, \mnras, 486, 4479. doi:10.1093/mnras/stz1155
\bibitem[L{\"u} et al.(2017)]{2017ApJ...835..181L} L{\"u}, H.-J., Zhang, H.-M., Zhong, S.-Q., et al.\ 2017, \apj, 835, 181. doi:10.3847/1538-4357/835/2/181
\bibitem[L{\"u} et al.(2022)]{2022ApJ...931L..23L} L{\"u}, H.-J., Yuan, H.-Y., Yi, T.-F., et al.\ 2022, \apjl, 931, L23. doi:10.3847/2041-8213/ac6e3a
\bibitem[Margalit et al.(2020)]{2020MNRAS.494.4627M} Margalit, B., Metzger, B.~D., \& Sironi, L.\ 2020, \mnras, 494, 4627. doi:10.1093/mnras/staa1036
\bibitem[Margalit \& Metzger(2017)]{2017ApJ...850L..19M} Margalit, B. \& Metzger, B.~D.\ 2017, \apjl, 850, L19. doi:10.3847/2041-8213/aa991c
\bibitem[McWilliams \& Levin(2011)]{2011ApJ...742...90M} McWilliams, S.~T. \& Levin, J.\ 2011, \apj, 742, 90. doi:10.1088/0004-637X/742/2/90
\bibitem[Metzger et al.(2010)]{2010MNRAS.406.2650M} Metzger, B.~D., Mart{\'\i}nez-Pinedo, G., Darbha, S., et al.\ 2010, \mnras, 406, 2650. doi:10.1111/j.1365-2966.2010.16864.x
\bibitem[Metzger et al.(2019)]{2019MNRAS.485.4091M} Metzger, B.~D., Margalit, B., \& Sironi, L.\ 2019, \mnras, 485, 4091. doi:10.1093/mnras/stz700
\bibitem[Metzger(2019)]{2019LRR....23....1M} Metzger, B.~D.\ 2019, Living Reviews in Relativity, 23, 1. doi:10.1007/s41114-019-0024-0
\bibitem[Mingarelli et al.(2015)]{2015ApJ...814L..20M} Mingarelli, C.~M.~F., Levin, J., \& Lazio, T.~J.~W.\ 2015, \apjl, 814, L20. doi:10.1088/2041-8205/814/2/L20
\bibitem[Most \& Philippov(2023a)]{2023ApJ...956L..33M} Most, E.~R. \& Philippov, A.~A.\ 2023a, \apjl, 956, L33. doi:10.3847/2041-8213/acfdae
\bibitem[Most \& Philippov(2023b)]{2023PhRvL.130x5201M} Most, E.~R. \& Philippov, A.~A.\ 2023b, \prl, 130, 245201. doi:10.1103/PhysRevLett.130.245201
\bibitem[Musolino et al.(2024)]{2024ApJ...962...61M} Musolino, C., Ecker, C., \& Rezzolla, L.\ 2024, \apj, 962, 61. doi:10.3847/1538-4357/ad1758
\bibitem[Nakar(2020)]{2020PhR...886....1N} Nakar, E.\ 2020, \physrep, 886, 1. doi:10.1016/j.physrep.2020.08.008
\bibitem[Neill et al.(2022)]{2022MNRAS.514.5385N} Neill, D., Tsang, D., van Eerten, H., et al.\ 2022, \mnras, 514, 5385. doi:10.1093/mnras/stac1645
\bibitem[Pannarale et al.(2015)]{2015PhRvD..92h1504P} Pannarale, F., Berti, E., Kyutoku, K., et al.\ 2015, \prd, 92, 081504. doi:10.1103/PhysRevD.92.081504
\bibitem[Peters(1964)]{1964PhRv..136.1224P} Peters, P.~C.\ 1964, Physical Review, 136, 1224. doi:10.1103/PhysRev.136.B1224
\bibitem[Pian et al.(2017)]{2017Natur.551...67P} Pian, E., D'Avanzo, P., Benetti, S., et al.\ 2017, \nat, 551, 67. doi:10.1038/nature24298
\bibitem[Piro et al.(2017)]{2017ApJ...844L..19P} Piro, A.~L., Giacomazzo, B., \& Perna, R.\ 2017, \apjl, 844, L19. doi:10.3847/2041-8213/aa7f2f
\bibitem[Pozanenko et al.(2018)]{2018ApJ...852L..30P} Pozanenko, A.~S., Barkov, M.~V., Minaev, P.~Y., et al.\ 2018, \apjl, 852, L30. doi:10.3847/2041-8213/aaa2f6
\bibitem[Radice et al.(2018a)]{2018ApJ...852L..29R} Radice, D., Perego, A., Zappa, F., et al.\ 2018a, \apjl, 852, L29. doi:10.3847/2041-8213/aaa402
\bibitem[Radice et al.(2018b)]{2018ApJ...869..130R} Radice, D., Perego, A., Hotokezaka, K., et al.\ 2018b, \apj, 869, 130. doi:10.3847/1538-4357/aaf054
\bibitem[Rezzolla et al.(2011)]{2011ApJ...732L...6R} Rezzolla, L., Giacomazzo, B., Baiotti, L., et al.\ 2011, \apjl, 732, L6. doi:10.1088/2041-8205/732/1/L6
\bibitem[Rhoades \& Ruffini(1974)]{1974PhRvL..32..324R} Rhoades, C.~E. \& Ruffini, R.\ 1974, \prl, 32, 324. doi:10.1103/PhysRevLett.32.324
\bibitem[Romani et al.(2022)]{2022ApJ...934L..17R} Romani, R.~W., Kandel, D., Filippenko, A.~V., et al.\ 2022, \apjl, 934, L17. doi:10.3847/2041-8213/ac8007
\bibitem[Rosswog et al.(2013)]{2013MNRAS.430.2585R} Rosswog, S., Piran, T., \& Nakar, E.\ 2013, \mnras, 430, 2585. doi:10.1093/mnras/sts708
\bibitem[Ruiz et al.(2018)]{2018PhRvD..98l3017R} Ruiz, M., Shapiro, S.~L., \& Tsokaros, A.\ 2018, \prd, 98, 123017. doi:10.1103/PhysRevD.98.123017
\bibitem[Ryu et al.(2010)]{2010PhRvC..82b5804R} Ryu, C.~Y., Kim, K.~S., \& Cheoun, M.-K.\ 2010, \prc, 82, 025804. doi:10.1103/PhysRevC.82.025804
\bibitem[Sasaki et al.(2016)]{2016PhRvL.117f1101S} Sasaki, M., Suyama, T., Tanaka, T., et al.\ 2016, \prl, 117, 061101. doi:10.1103/PhysRevLett.117.061101
\bibitem[Savchenko et al.(2017)]{2017ApJ...848L..15S} Savchenko, V., Ferrigno, C., Kuulkers, E., et al.\ 2017, \apjl, 848, L15. doi:10.3847/2041-8213/aa8f94
\bibitem[Schianchi et al.(2024)]{2024PhRvD.109l3011S} Schianchi, F., Ujevic, M., Neuweiler, A., et al.\ 2024, \prd, 109, 123011. doi:10.1103/PhysRevD.109.123011
\bibitem[Sridhar et al.(2021)]{2021MNRAS.501.3184S} Sridhar, N., Zrake, J., Metzger, B.~D., et al.\ 2021, \mnras, 501, 3184. doi:10.1093/mnras/staa3794
\bibitem[Tan et al.(2020)]{2020PhRvL.125z1104T} Tan, H., Noronha-Hostler, J., \& Yunes, N.\ 2020, \prl, 125, 261104. doi:10.1103/PhysRevLett.125.261104
\bibitem[Tanaka et al.(2017)]{2017PASJ...69..102T} Tanaka, M., Utsumi, Y., Mazzali, P.~A., et al.\ 2017, \pasj, 69, 102. doi:10.1093/pasj/psx121
\bibitem[Tews et al.(2021)]{2021ApJ...908L...1T} Tews, I., Pang, P.~T.~H., Dietrich, T., et al.\ 2021, \apjl, 908, L1. doi:10.3847/2041-8213/abdaae
\bibitem[Troja et al.(2016)]{2016ApJ...827..102T} Troja, E., Sakamoto, T., Cenko, S.~B., et al.\ 2016, \apj, 827, 102. doi:10.3847/0004-637X/827/2/102
\bibitem[Troja(2023)]{2023Univ....9..245T} Troja, E.\ 2023, Universe, 9, 245. doi:10.3390/universe9060245
\bibitem[Tsang et al.(2012)]{2012PhRvL.108a1102T} Tsang, D., Read, J.~S., Hinderer, T., et al.\ 2012, \prl, 108, 011102. doi:10.1103/PhysRevLett.108.011102
\bibitem[Watson et al.(2019)]{2019Natur.574..497W} Watson, D., Hansen, C.~J., Selsing, J., et al.\ 2019, \nat, 574, 497. doi:10.1038/s41586-019-1676-3
\bibitem[Wei et al.(2016)]{2016arXiv161006892W} Wei, J., Cordier, B., Antier, S., et al.\ 2016, arXiv:1610.06892. doi:10.48550/arXiv.1610.06892
\bibitem[Yang et al.(2018)]{2018ApJ...856..110Y} Yang, H., East, W.~E., \& Lehner, L.\ 2018, \apj, 856, 110. doi:10.3847/1538-4357/aab2b0
\bibitem[Yuan et al.(2023)]{2023PhRvD.108h3018Y} Yuan, H.-Y., L{\"u}, H.-J., Rice, J., et al.\ 2023, \prd, 108, 083018. doi:10.1103/PhysRevD.108.083018
\bibitem[Yuan et al.(2021)]{2021ApJ...912...14Y} Yuan, Y., L{\"u}, H.-J., Yuan, H.-Y., et al.\ 2021, \apj, 912, 14. doi:10.3847/1538-4357/abedb1
\bibitem[Zhang et al.(2018)]{2018NatCo...9..447Z} Zhang, B.-B., Zhang, B., Sun, H., et al.\ 2018, Nature Communications, 9, 447. doi:10.1038/s41467-018-02847-3
\bibitem[Zhang(2016)]{2016ApJ...827L..31Z} Zhang, B.\ 2016, \apjl, 827, L31. doi:10.3847/2041-8205/827/2/L31
\bibitem[Zhang(2018)]{2018pgrb.book.....Z} Zhang, B.\ 2018, The Physics of Gamma-Ray Bursts by Bing Zhang. ISBN: 978-1-139-22653-0. Cambridge Univeristy Press, 2018. doi:10.1017/9781139226530
\bibitem[Zuraiq et al.(2024)]{2024PhRvD.109b3027Z} Zuraiq, Z., Mukhopadhyay, B., \& Weber, F.\ 2024, \prd, 109, 023027. doi:10.1103/PhysRevD.109.023027




       


\end{thebibliography}
\end{document}